%% file: main.tex
\documentclass[%
superscriptaddress,
groupedaddress,
preprint,
showpacs,preprintnumbers,
 amsmath,amssymb,
prstab,
]{revtex4-1}

\usepackage{graphicx}% Include figure files
\usepackage{dcolumn}% Align table columns on decimal point
\usepackage{bm}% bold math
\usepackage{subfig}
\usepackage{tikz}
\usepackage{color}
\usepackage{caption, booktabs}
\usepackage[arrow,matrix]{xy}
\usepackage{minibox}

\input{TEDmakros}

\begin{document}

%\preprint{APS/123-QED}

\title{Beam Dynamics Analysis of Dielectric Laser Acceleration using a Fast 6D Tracking Scheme}
%\title{A Fast 6D Particle Tracking Scheme for the Design of Dielectric Laser Acceleration Structures}% Force line breaks with \\

\author{Uwe Niedermayer}
 \email{niedermayer@temf.tu-darmstadt.de}
 \affiliation{%
Institut f\"ur Theorie elektromagnetischer Felder, Technische Universit\"at Darmstadt, Schlossgartenstr. 8 D-64289 Darmstadt, Germany
}%
\author{Thilo Egenolf}%
 \affiliation{%
Institut f\"ur Theorie elektromagnetischer Felder, Technische Universit\"at Darmstadt, Schlossgartenstr. 8 D-64289 Darmstadt, Germany
}%
\author{Oliver Boine-Frankenheim}
 \altaffiliation[Also at ]{GSI Helmholtzzentrum f\"ur Schwerionenforschung, Planckstr. 1, D-64291 Darmstadt, Germany}
 \affiliation{%
Institut f\"ur Theorie elektromagnetischer Felder, Technische Universit\"at Darmstadt, Schlossgartenstr. 8 D-64289 Darmstadt, Germany
}%

\date{\today}% It is always \today, today,

\begin{abstract}
A six-dimensional symplectic tracking approach exploiting the periodicity properties of Dielectric Laser Acceleration (DLA) gratings is presented. The longitudinal kick is obtained from the spatial Fourier harmonics of the laser field within the structure, and the transverse kicks are obtained using the Panofsky-Wenzel theorem. Additionally to the usual, strictly longitudinally periodic gratings, our approach is also applicable to periodicity chirped (sub-relativistic) and tilted (deflection) gratings. In the limit of small kicks and short periods we obtain the 6D Hamiltonian, which allows, for example, to obtain matched beam distributions in DLAs. The scheme is applied to beam and grating parameters similar to recently performed experiments. The paper concludes with an outlook to laser based focusing schemes, which are promising to overcome fundamental interaction length limitations, in order to build an entire microchip-sized laser driven accelerator.
%\begin{description}
%\end{description}
\end{abstract}

%\pacs{Valid PACS appear here}
\maketitle

\section{Introduction}
Dielectric Laser Acceleration (DLA) provides highest gradients among non-plasma accelerators. In 2013, the acceleration of relativistic electrons was demonstrated at SLAC with a gradient of more than 250~MeV/m~\cite{Peralta2013DemonstrationMicrostructure.}, which was recently increased to 690~MeV/m~\cite{Wootton2016DemonstrationPulses}. 
Low energy electrons (27.7~keV) were accelerated by the group in Erlangen~\cite{Breuer2013Laser-BasedStructure} with a gradient of 25~MeV/m using a single grating structure. The group at Stanford University used a dual pillar structure to accelerate 96~keV electrons with a gradient of more than 200~MeV/m~\cite{Leedle2015LaserStructure}.
In principle, the gradient of a DLA is only limited by the structure damage threshold fluence, which is roughly two orders of magnitude higher for dielectrics than for metals. 
The reason for the rediscovery of this rather old concept of inverse Smith-Purcell or inverse Cherenkov effects for particle acceleration (see e.g.~\cite{Shimoda1962ProposalMaser}) is that nowadays both the ultrashort laser pulse control techniques as well as the nano-fabrication have significantly improved. Summaries of the recent developments can be found in~\cite{England2014DielectricAccelerators} and~\cite{Wootton2016DielectricApplications}.

Although the experimentally demonstrated gradients in DLA structures are very promising, there are still crucial challenges to create a miniaturized DLA-based particle accelerator. So far the experimentally achieved gradients could only be used to increase the beam's energy spread and not for coherent acceleration. Moreover, the interaction length with present DLA structures is limited to the Rayleigh range (see App.~\ref{Rayleigh}) of the incident electron beam. For low energy electrons, due to the high gradient, the acceleration defocusing even leads to interaction distances significantly shorter than the Rayleigh range. 

\begin{figure}[h]
				\centering
	\includegraphics[width=0.5\textwidth]{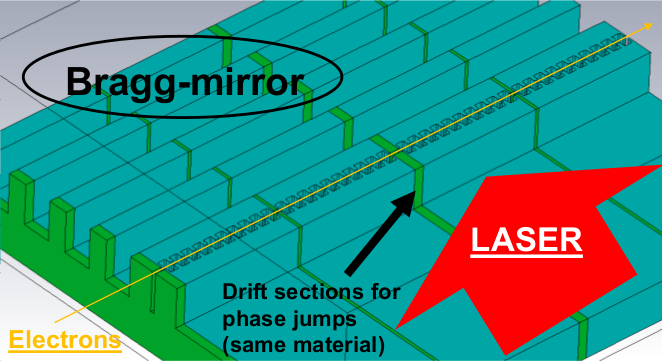}
    \caption{Example Bragg cavity grating with a Bragg mirror on one side. }
    \label{APF}
\end{figure}

Thus, in order to use DLA for a real accelerator, focusing schemes have to be developed. One option would be Alternating Phase Focusing (APF) as outlined in Fig.~\ref{APF}. 
Here, drift sections between grating cells lead to jumps in the synchronous phase, which can be designed to provide net focusing. Such schemes can be a way to increase the interaction length in DLAs and make an accelerator on a microchip feasible.

A challenge in the creation of a DLA based optical accelerator is related to the complex 3D beam dynamics in DLA structures, which has not been treated systematically in the existing literature yet.   
In order to facilitate front to end simulations and identify optimized DLA structures, 
we employ a simple and efficient numerical tracking scheme, which does not require a large amount of computing power, it runs in Matlab~\cite{MathWorks2016Matlab} on an ordinary PC.

Due to the periodicity of a DLA grating, only one spatial Fourier harmonic contributes to the kick of the beam. In a simplified approach, where fringe fields are neglected, the entire laser field can be represented by a set of such Fourier coefficients, where only one complex number represents each grating cell. By means of the Panofsky-Wenzel theorem~\cite{Panofsky1956SomeFields}, this single complex number also allows to determine the transverse kicks experienced by a particle while traveling through one grating period.

When the three-dimensional kicks are applied to the beam particles in a symplectic scheme (we use symplectic Euler, which is equivalent to Leap Frog) the tracking becomes phase space volume preserving. Thus, with no numerical (artificial) emittance increase, the physical emittance increase due to the nonlinear fields in the DLA interaction can be calculated. Moreover, since the equations of motion are coupled, there is also emittance exchange between the different planes which can be analyzed.

In the present study we neglect all intensity dependent effects as space charge, wakes, and radiation emission. The number of particles is chosen such that smooth spectra are obtained, a reasonable value is $10^6$, at which the computational time is about one second per grating cell.

With no loss of generality, we restrict ourselves to symmetric grating structures driven from both lateral sides. This makes sure that the axis of symmetry is in the center and the fields have a $\cosh$ profile. In the case of non-symmetric structures or non-symmetric driving the fields will have an exponential or an off-axis $\cosh$ profile. However, single driver systems can be combined with Bragg mirrors in order to obtain a good approximation to an on-axis $\cosh$ profile with a single side driver (see again Fig.~\ref{APF}).

As it is usually done e.g. for the synchrotron motion in ion synchrotrons, we take the limit from the tracking difference equations to differential equations. Since the three-dimensional kick must be irrotational due to the Panofsky-Wenzel theorem, it can be derived from a scalar potential. This potential directly allows to determine the 6D Hamiltonian which completely describes the single particle dynamics analytically.

\begin{figure}[t]
				\centering
	\includegraphics[width=0.5\textwidth]{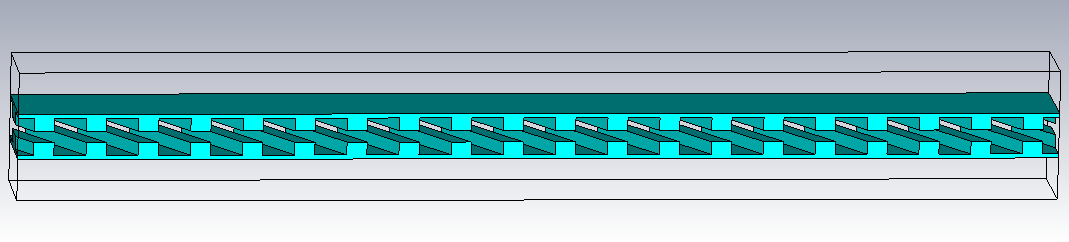}
    \caption{Tilted Bragg cavity grating, where the dual drive laser comes from top and bottom and is polarized in the electron beam direction (left to right).}
    \label{LongTilted}
\end{figure}
Both the numerical and the analytical approach can be generalized from ordinary DLA gratings to tilted DLA gratings, which have been proposed as deflectors or laser driven undulators~\cite{Plettner2008ProposedUndulator,Plettner2008Microstructure-basedLaser,Plettner2009Photonic-basedStructures}. Such a grating is depicted in Fig.~\ref{LongTilted}. However, since our code does not include the radiation fields, a dedicated code as e.g.~\cite{Fallahi2016MITHRALasers} can be used to treat the dynamics self consistently. The analytical kicks reported here can serve as input quantities. Our approach aims at maximal simplicity such that studies of fundamental questions, as e.g. transverse focusing and deflection, are quickly possible.

The paper is organized as follows. Section~\ref{Periodic} presents the determination of the longitudinal and transverse fields and kicks in a single grating period. Here we use CST Studio Suite~\cite{CST2016CSTSuite} to calculate the longitudinal kick at the center of the structure. The dependence on the transverse coordinates as well as the transverse kicks are modeled analytically. In Sect.~\ref{Tracking} we present a symplectic 6D tracking method based on one kick per grating period.
Analytical descriptions of the coupled longitudinal and transverse beam dynamics as well as the full 6D Hamiltonian are given in Sect.~\ref{Analytic}. Simplifications and beam matching in linearized fields are also discussed in this section.
In Sect.~\ref{Applications} we address the three crucial examples: subrelativistic acceleration, relativistic acceleration, and deflection by means of DLA gratings. 
The paper concludes with a summary and an outlook to DLA focusing channels in Sect.~\ref{Conclusion}.

\section{Fields and kicks in periodic structures}
\label{Periodic}
\begin{figure}[h]
				\centering
	\includegraphics[width=0.5\textwidth]{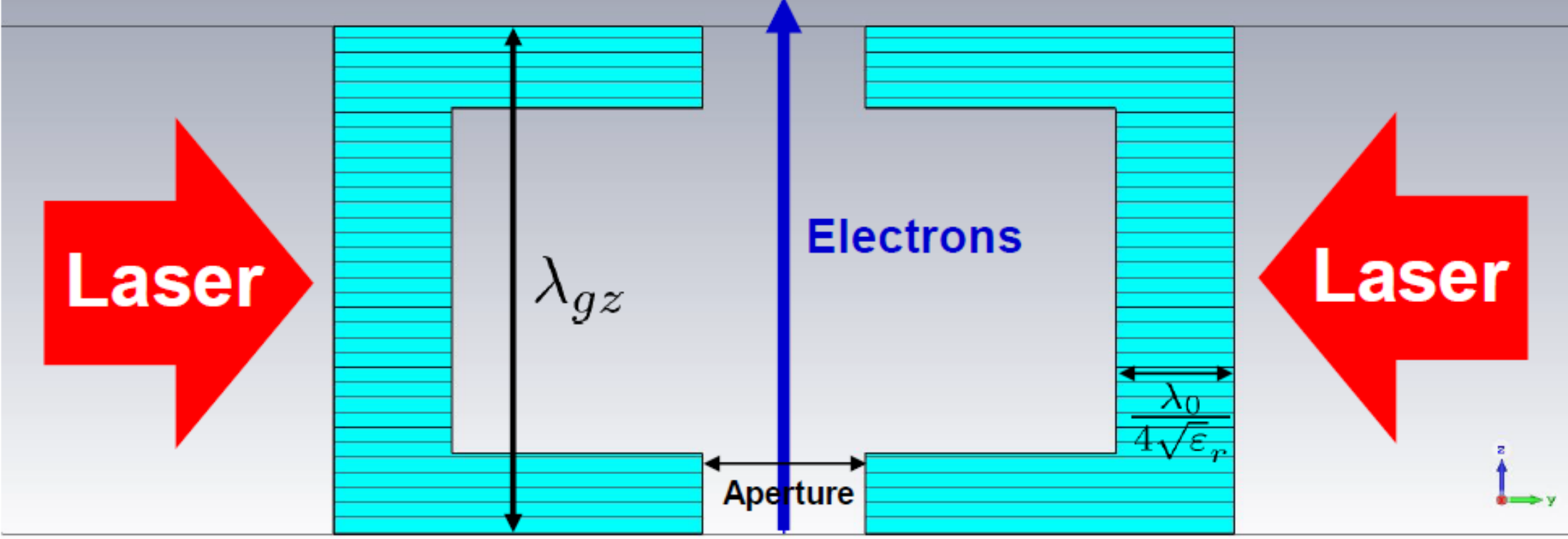}
    \caption{One period of a symmetric Bragg mirror cavity structure.}
    \label{Cavity}
\end{figure}

Usual particle tracking algorithms solve Maxwell's equations with a predefined time step.
Instead of that,
we make use of the periodicity of the structure and apply only the kicks which are known not to average out a-priori. The other field harmonics are neglected. The validity of this neglect depends on the effect of transients which is effectively suppressed when the structure period is matched to the beam velocity.  
With no loss of generality we restrict ourselves here to an infrared laser with $\lambda_0=1.96\;\mu$m and structures made of Silicon ($\eps_r=11.63$). 
A single cell of a symmetrically driven Bragg mirror cavity structure is shown in Fig.~\ref{Cavity}.

\subsection{Analysis of the longitudinal field}
A coordinate system is applied such that the electron beam propagates in positive z-direction and the z-polarized laser propagates in y-direction.
The unit cell of a periodic dielectric structure has dimensions $\lambda_{gx}$ and $\lambda_{gz}$. In order to allow the laser field to escape the structure, open boundaries in positive and negative y-direction are assumed. 
\begin{figure}[b]
				\centering
	\includegraphics[width=0.45\textwidth]{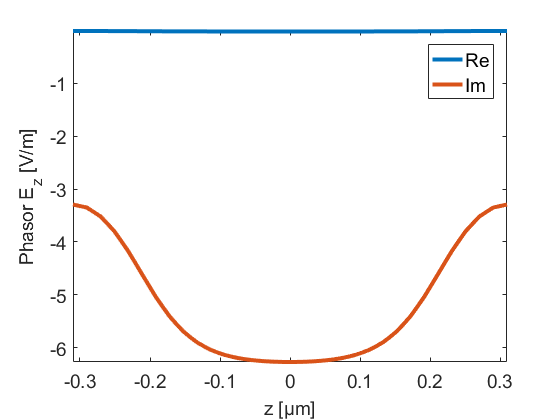}
    \includegraphics[width=0.45\textwidth]{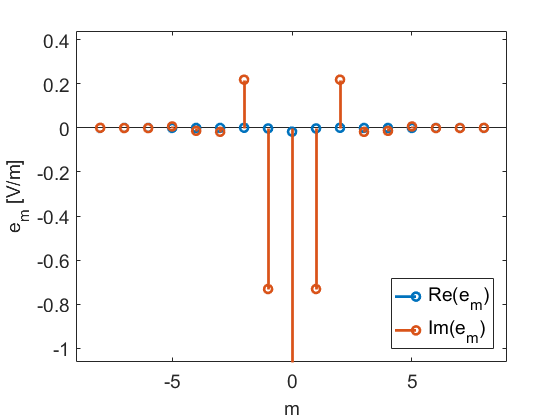}
    \caption{Longitudinal electric field on the beam axis for the Bragg mirror structure in Fig.~\ref{Cavity} and spatial Fourier harmonics for incident laser field normalized to 1~V/m.}
    \label{FieldOnAxis}
\end{figure}
The energy gain of a particle in one cell is
\begin{align}
\Delta W(x,y;s)&= q\int\limits_{-\lambda_{gz}/2}^{\lambda_{gz}/2} E_{z} (x,y,z; t=(z+s)/v) \dz  \\
																 &= q\int\limits_{-\lambda_{gz}/2}^{\lambda_{gz}/2} \Re\{\Ec_{z} (x,y,z) e^{i\omega (z+s)/v} \}\dz,  
\label{Egain}
\end{align}
where the underlined electric field is a phasor at the fixed frequency $\omega=2\pi c/\lambda_0$ of the laser, and $q$ is the charge ($q=-e$ for electrons). The variable $s$ denotes the relative position of the particle behind an arbitrarily defined reference particle moving at $z=vt$. Thus $z$ is the absolute position in the laboratory frame, while $s$ denotes the phase shift w.r.t. $z$.
Due to the $z$-periodicity, the laser field can be expanded in spatial Fourier series 
\begin{align}
\Ec_{z} (x,y,z)  &=\sum\limits_{m=-\infty}^\infty \ec_m(x,y) e^{-im\frac{2\pi}{\lambda_{gz}}z} \\
\ec_m(x,y)&=\frac{1}{\lambda_{gz}} \int\limits_{-\lambda_{gz}/2}^{\lambda_{gz}/2} \Ec_{z} (x,y,z) e^{im\frac{2\pi}{\lambda_{gz}}z} \dz,
\label{fouriercoefficient}
\end{align}
which allows to compute the energy gain integral (Eq.~\ref{Egain}) as
\beq
\Delta W(x,y;s)=q\Re\left\{e^{2\pi i \frac{s}{\beta\lambda_0}}
\sum\limits_{m=-\infty}^\infty \ec_m(x,y) \lambda_{gz} \mathrm{sinc}
\left[\pi\left(\frac{\lambda_{gz}}{\beta\lambda_0}-m\right)\right]  \right\}.  
\label{sinc}
\eeq
The electric field phasor and its spatial Fourier coefficients for the structure in Fig.~\ref{Cavity} are plotted in Fig.~\ref{FieldOnAxis}.
It has a small real part, which is coincidental, and a strong first and weak second harmonic.
If the round braces in Eq.~\ref{sinc} is non-integer, the energy gain averages to zero, if it is integer other than zero, it directly vanishes.
Thus we have the phase synchronicity condition
\beq
\lambda_{gz}=m\beta\lambda_0
\label{phasesynchronicity}
\eeq
and the particle's energy gain simplifies to
\beq
\Delta W(x,y;s)=q\lambda_{gz}\Re\left\{e^{2\pi i \frac{s}{\beta\lambda_0}} \ec_{m}(x,y)\right\} = q\lambda_{gz} |\ec_{m}|\cos\left(2\pi \frac{s}{\beta\lambda_0} +\phi_m \right),
\label{SimpleEnergyGain}
\eeq
where $\phi_m=\arctan \Im\{\ec_{m}\}/\Re\{\ec_{m}\}$ is the phase of the Fourier coefficient. The energy gain is maximal at $s_{opt}=-\phi_m \beta \lambda_0/2\pi$ for positively charged particles and $s_{opt}=-(\phi_m+\pi) \beta \lambda_0/2\pi$ for electrons. Zero acceleration is found at $s_{0}=-(\phi_m \pm \pi/2) \beta \lambda_0/2\pi$, where all these expressions are to be taken modulo $\beta \lambda_0/m$.
The integer spatial harmonic $m$ has the same meaning as the harmonic number in conventional accelerators, i.e. the number of buckets per grating period and Eq.~\ref{phasesynchronicity} resembles the Wideroe condition. The non-synchronous harmonics, which we assume to average out in this framework, however provide kicks in the case $s$ in Eq.~\ref{sinc} is not a constant. This leads then to multi-harmonic buckets and can also be used for ponderomotive focusing, see~\cite{Naranjo2012StableHarmonics}.

For sub-relativistic accelerators, the grating needs to be chirped in period length in order to always fulfill Eq.~\ref{phasesynchronicity} on the energy ramp. The change of period length is given by the energy velocity differential
\begin{equation}
\frac{\Delta z}{\lambda_{gz}}=\frac{1}{\beta^2\gamma^2}\frac{\Delta W}{W}
\end{equation}
and is in the range of $\lesssim 1\%$ for $W_\mathrm{kin}=30$ keV and $\Delta W/\lambda_{gz}=1$ GeV/m. The thus created "quasi-periodic" gratings can be seen in good approximation as periodic, however, phase drifts have to be compensated in the structure design~\cite{Niedermayer2017DesigningChip}.

\subsection{Analysis of the transverse field}
\begin{figure}[t]
    \centering
		\includegraphics[width=0.65\textwidth]{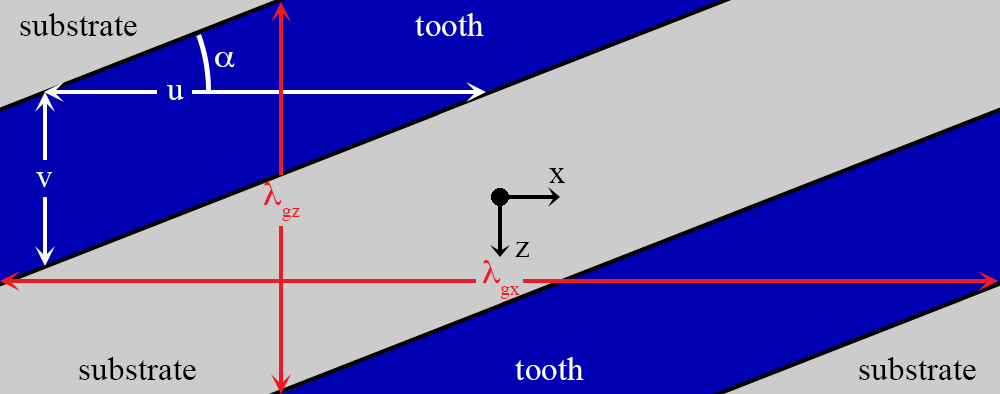}
    \caption{Tilted grating with periodic boundary conditions in $x$ and $z$ direction and $\tan\alpha=\lambda_\mathrm{gz}/\lambda_\mathrm{gx}=v/u$. 
    %The origin of the coordinate system is in the center.
    }
    \label{tilted}
\end{figure}

The transverse field probed by a rigidly moving charge can be obtained using the Panofsky-Wenzel theorem~\cite{Panofsky1956SomeFields}, which holds for either vanishing fields at infinity or periodic boundary conditions as
\beq
\nabla'\times\Delta\pv(\rv_\perp,s) = \int\limits_{-T/2}^{T/2}\dt \left[ \nabla\times\Fv (\rv_\perp,z,t)\right]_{z=vt-s}=\left. \Bv\right|_{-T/2}^{T/2}=0.
\label{PWtheorem}
\eeq
Here the 'relative gradient' is defined as $\nabla'=(\partial_x,\partial_y,-\partial_s)^\mathrm{T}$ and $\lambda_{gz}=\beta cT$.
The transverse kick per cell can be written as 
\begin{align}
\Delta\pv_\perp(x,y;s)&=-\int\ds\nabla\!_\perp \Delta p_\parallel(x,y;s) \\
&=-\frac{\lambda_{gz}}{2\pi m} q\frac{1}{\beta c}\nabla\!_\perp \int\limits_{-\lambda_{gz}/2}^{\lambda_{gz}/2} \Im\{\Ec_{z} (x,y,z) e^{i\omega (z+s)/v} \}\dz,  
\end{align}
where the energy momentum differential $\Delta p_\parallel =\Delta W /(\beta c)$ was applied.
Moreover, if the phase-synchronicity condition (Eq.~\ref{phasesynchronicity}) is fulfilled, the kick becomes
\begin{align}
\Delta\pv_\perp(x,y;s)&=-\frac{\lambda_{gz}^2}{2\pi m} q\frac{1}{\beta c}\nabla\!_\perp \Im\left\{e^{2\pi i \frac{s}{\beta\lambda_0}} \ec_{m}(x,y)\right\}\\
											&=-\frac{\lambda_{gz}}{m}q\frac{1}{\beta c} \Im\left\{e^{2\pi i \frac{s}{\beta\lambda_0}} \fvc_{m}(x,y)\right\},
\label{transversekick}
\end{align}
where $\fvc_m(x,y)=\lambda_{gz}\nabla\!_\perp \ec_m(x,y)/2\pi$.
In the following, the structure under investigation is generalized to a tilted grating as visible in Fig.~\ref{tilted}, which reproduces the ordinary grating for tilt angle $\alpha=0$.
The tilted grating is periodic in $z$ and $x$ direction. Thus for any function $F(x,y,s)$ must hold
\beq
\frac{\partial F}{\partial x} =\frac{\partial F}{\partial s} \frac{\partial s}{\partial z} \frac{\partial z}{\partial x} = \frac{\partial F}{\partial s} \tan \alpha.
\eeq
The $s$-derivative can be calculated in the Fourier representation by Eq.~\ref{SimpleEnergyGain} as
\beq
\frac{\partial\ec_m (x,y)}{\partial x}=\tan \alpha \frac{2\pi i}{\beta \lambda_0} \ec_m (x,y).
\eeq
The derivatives in $y$-direction can be determined by the dispersion relation for the synchronous mode. We have 
\beq
k_z=\frac{\omega}{\beta c},\;\;\; k_x=\frac{2\pi }{\beta \lambda_0} \tan\alpha \;\;\; \mathrm{and} \;\;\; k=\frac{\omega}{c} 
\eeq
and thus 
\beq
k_y=\pm \sqrt{k^2-(k_z^2+k_x^2)}=\pm \frac{\omega}{c}\sqrt{ 1-\frac{1}{\beta^2}(1+\tan^2 \alpha)}.
\label{ky}
\eeq
For a non-tilt grating ($\alpha=0$) this is the well known evanescent decay of the near field $k_y=i\omega/(\beta\gamma c)$. 
Once $k_x,k_y$ are determined, the fields can be found from
\beq
\ec_m(x,y)=\ec_m(0,0)\cosh(ik_y y)e^{i k_x x},
\label{AnaTiltGrating}
\eeq
where $\lambda_{gx}=\lambda_{gz}/\tan\alpha$.

\begin{figure}[t]
    \centering
		\includegraphics[width=\textwidth]{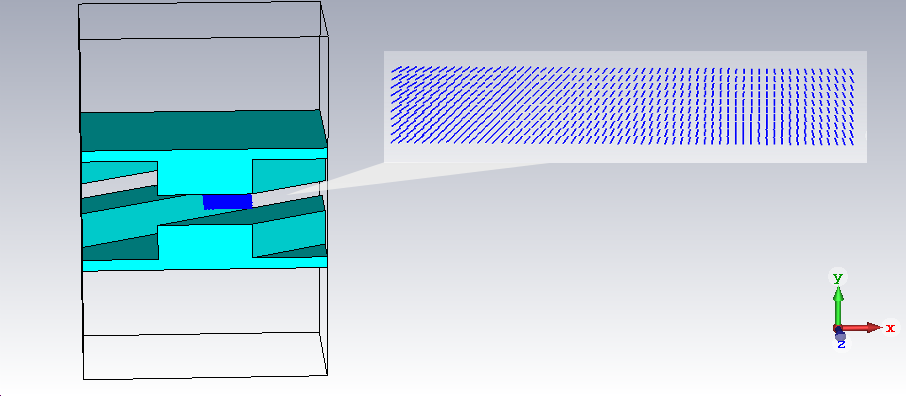}
\caption{Single cell of the tilted grating deflector structure and enlargement of kick integration curves array. The integrated kicks are displayed in Fig.~\ref{Kicks}.}
    \label{tilted3D}
\end{figure}

\begin{figure}[t]
				\centering
    \includegraphics[width=\textwidth]{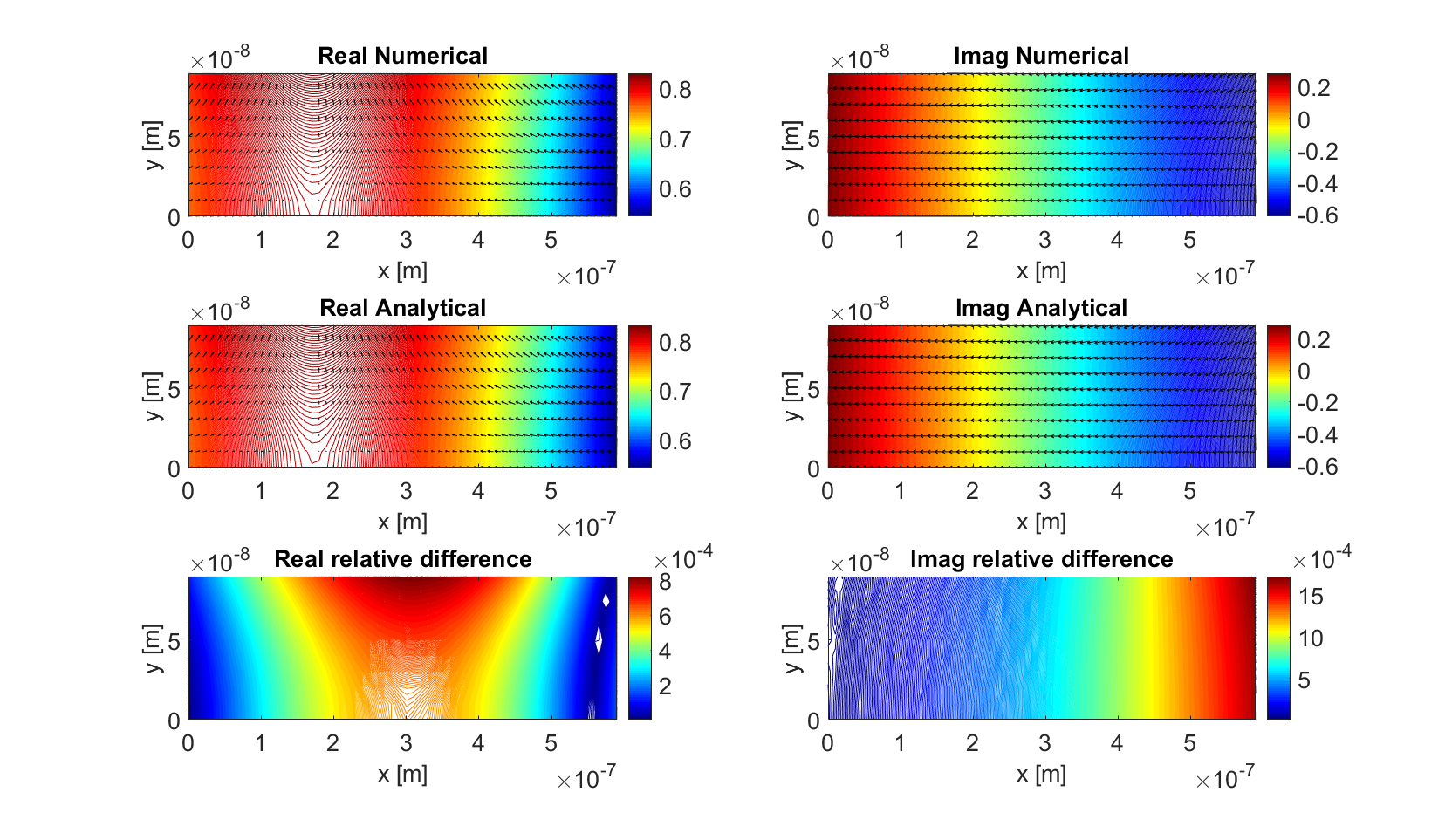}
    \caption{Contour lines of $\ec_1(x,y)$ and kick field $\fvc_1(x,y)$ for the tilted grating with $\alpha=30\deg$. The fields have been obtained both analytically and numerically, the bottom plots show the relative difference.}
    \label{Kicks}
\end{figure}

A map of the energy gain and transverse kicks for the grating in Fig.~\ref{tilted3D} can be seen in Fig.~\ref{Kicks} for a grating tilt angle $\alpha=30\deg$.
The results labeled numerical are obtained by line integration (Eq.~\ref{SimpleEnergyGain}) of the electric field simulated with CST MWS~\cite{CST2016CSTSuite} and the analytical results correspond to Eq.~\ref{AnaTiltGrating}.
The transverse kicks are obtained by Eq.~\ref{transversekick} as 
\begin{align}
\fvc_m(x,y)=\; &\ec_m(0,0)\cosh(ik_y y)e^{i k_x x} i m \tan \alpha \ex \nonumber\\
			 +&\ec_m(0,0)\sinh(ik_y y)e^{i k_x x} (ik_y \lambda_{gz}/2\pi)\ey
             \label{transkick}
\end{align}
and are depicted as arrows in Fig.~\ref{Kicks}. For the numerical results, the gradient is determined by finite differences in Matlab~\cite{MathWorks2016Matlab}.
Note that $-ik_y \in \mathbb{R}^+$, i.e. the kick in $x$-direction is in phase with the acceleration while the kick in $y$-direction is $90$ degrees shifted.

For a particle that is only slightly displaced from the beam axis by $\Delta \vec x=(\Delta x, \Delta y)$ the kick can be written as two-dimensional Taylor expansion
\begin{align}
\fvc_m(x,y)&=\fvc_m(x_0,y_0)+(\nabla\!_\perp \fvc_m(x_0,y_0))\Delta\vec x +\mathcal O (||\Delta\vec x||^2) \nonumber\\
		  &=\frac{\lambda_{gz}}{2\pi} \left(\nabla\!_\perp \ec_m(x_0,y_0)  +(\nabla\!_\perp\nabla\!_\perp^\mathrm{T}) \ec_m(x_0,y_0)\Delta\vec x  \right) +\mathcal O (||\Delta\vec x||^2),
					\label{transverseExpansion}
\end{align}
where 
\beq
\nabla\!_\perp\nabla\!_\perp^\mathrm{T} =
\begin{pmatrix}
\partial_x^2 & \partial_x\partial_y \\
\partial_y\partial_x & \partial_y^2
\end{pmatrix}
\eeq
is the Hessian.
The expansion Eq.~\ref{transverseExpansion} about $x_0=0, y_0=0$ of Eq.~\ref{AnaTiltGrating} results in  
\beq
\fvc_m(\Delta x, \Delta y)= \dfrac{\lambda_{gz}}{2\pi}\ec_m(0,0)
\begin{pmatrix}
i k_x - k_x^2 \Delta x  \\
-k_y^2 \Delta y
\end{pmatrix},
\eeq 
i.e. a position independent (coherent) kick component in $x$-direction, vanishing for ${\alpha=0}$. Using this abstract derivation, the results of several papers proposing DLA undulators~\cite{Plettner2008ProposedUndulator,Plettner2008Microstructure-basedLaser,Plettner2009Photonic-basedStructures} can be recovered.

\section{Tracking equations}
\label{Tracking}
In order to study the motion of particles in the fields of periodic gratings we approximate the forces by one kick per grating period and track with the symplectic Euler method. In spite of the very high gradients in DLA structures, the energy can still be seen as an adiabatic variable, as it is the case in conventional linacs. Tracking the full time dependence of $\gamma$, as required for example in plasma accelerators, can be avoided due to the shortness of the periods. 
For simplicity, we restrict ourselves to $m=1$ from this point and introduce normalized variables in the paraxial approximation
\beq
x'=\frac{p_x}{p_{z0}}\;,\;\;\; \Delta x'=\frac{\Delta p_x(x,y,\phi)}{p_{z0}}\;,\;\;\; y'=\frac{p_y}{p_{z0}}\;,\;\;\;\Delta y'=\frac{\Delta p_y(x,y,\phi)}{p_{z0}},   \nonumber
\eeq
\beq
\phi=2\pi \frac{s}{\lambda_{gz}}\;,\;\;\; \delta=\frac{W-W_0}{W_{0}}\;,\;\;\; \Delta\delta=\frac{\Delta W(x,y,\phi)-\Delta W(0,0,\phi_\mathrm{s})}{W_{0}}, 
\label{variables}
\eeq
where $W_0=\gamma m_e c^2$ and $p_{z0}=\beta\gamma m_ec$.
The particle at the synchronous phase $\phi_\mathrm{s}$ has $\Delta\delta=0$, i.e. its energy gain is entirely described by the acceleration ramp.
The energy gain $\Delta W$ is given by Eq.~\ref{SimpleEnergyGain} and thus the energy gain of the synchronous particle is 
\beq
\Delta W(0,0,\phi_\mathrm{s})=q \lambda_{gz}\Re\left\{e^{i\phi_\mathrm{s}} \ec_{1}\right\},
\eeq
where we write $\ec_1=\ec_1(x=0,y=0)$ for brevity. Note that the synchronous phase and the phase of each particle always refer to the laser phase. 
The sum of the kicks 
\begin{equation}
W(N)=W_\mathrm{init}+\sum_{n=1}^N\Delta W^{(n)}(0,0,\phi_\mathrm{s}^{(n)})
\end{equation}
describes the acceleration ramp, where the synchronous phase $\phi_\mathrm{s}$ can be chosen arbitrarily in each grating cell. 
The variables $e_1$, $\lambda_{gz}$, $W_0$, $\beta$, $\gamma$, $\phi_\mathrm{s}$ and all variables in Eq.~\ref{variables} are stored as arrays indexed by the grating cell number.
The kicks are obtained using Eqs.~\ref{SimpleEnergyGain}, \ref{AnaTiltGrating}, \ref{transkick}, and~\ref{transversekick} and read
\begin{subequations}
\begin{align}
\Delta x'&=-\frac{q\lambda_0}{p_{z0} c}\tan(\alpha)\cosh(ik_yy)   \Re\left\{\ec_1 e^{i\phi+i\frac{2\pi x}{\lambda_{gx}}}  \right\}  \\
\Delta y'&=\frac{-ik_y \lambda_0^2 q \beta}{2\pi p_{z0} c}\sinh(ik_yy) 
\Im\left\{\ec_1 e^{i\phi+i\frac{2\pi x}{\lambda_{gx}}} \right\}  \\
\Delta\delta&=\frac{q\lambda_{gz}}{\gamma m_e c^2} \Re \left\{ \ec_1 \left(
            	\cosh(ik_yy)
                e^{i\phi+i\frac{2\pi x}{\lambda_{gx}}}
                -e^{i\phi_s}\right)
            \right\},
\end{align}
\end{subequations}
where $k_y$ is given by Eq.~\ref{ky}.
The tracking equations are
\beq
\left(\begin{array}{c} 
x  \\ 
x' \\
y  \\ 
y' \\
\phi  \\ 
\delta \\
\end{array}\right)^{(n+1)}
=
\left(\begin{array}{c} 
x  \\ 
A x' +\Delta x'(x,y,\phi)\\
y  \\ 
A y' +\Delta y'(x,y,\phi)\\
\phi\\ 
\delta +\Delta \delta(x,y,\phi;\phi_\mathrm{sync})\\
\end{array}\right)^{(n)}
+
\left(\begin{array}{c} 
\lambda_{gz} x'(x,y,\phi)\\ 
0 \\
\lambda_{gz} y'(x,y,\phi)\\ 
0 \\
-\frac{2\pi}{\beta^2\gamma^2}\delta(x,y,\phi)\\ 
0 \\
\end{array}\right)^{(n+1)},
\label{tracking6D}
\eeq
where an explicit scheme is obtained by applying first the 'kicks' and then the 'pushes'.
The adiabatic damping in the transverse planes is described by 
\begin{equation}
A^{(n)}=\frac{(\beta\gamma)^{(n+1)}}{(\beta\gamma)^{(n)}}=1+
\left[\frac{\lambda_0 q\Re\left\{e^{i\phi_\mathrm{s}} \ec_{1}\right\} } {\beta \gamma m_e c^2} \right]^{(n)}.
\end{equation}
Symplecticity of the scheme is confirmed by calculating 
\beq
\mathrm{det}\frac{\partial(x,x',y,y',\phi,\delta)^{(n+1)}}{\partial(x,x',y,y',\phi,\delta)^{(n)}}={A^{(n)}}^2,
\eeq
which holds independently of the realization of the kick functions $\Delta x'(x,y,\phi)$, $\Delta y'(x,y,\phi)$ and $\Delta \delta(x,y,\phi)$.
Equations~\ref{tracking6D} contain the full non-linear kicks which cannot be linearized, since the usual bunch lengths and widths are not significantly smaller than the grating periods and apertures. However, in the idealized case of extremely small bunches, linearization results in a scheme equivalent to one linear R-matrix transformation per grating period.

Relevant information about the particle ensemble moving in space is given by statistical quantities such as envelope and emittance, which
can be derived from the beam matrix (second order moment matrix) as function of the period number. We define the 6D coordinate vector as
\begin{equation}
\rv=\left(
x,
p_x,
y,
p_y,
\Delta s,
\Delta P_z
\right)^T,
\end{equation}
where $\Delta s=(\phi-\phi_s)\lambda_{gz}/2\pi$ and $\Delta P_z=\Delta p_z/\gamma=W_0/(\beta c \gamma) \delta$.
The symmetric and positive definite beam matrix reads
\begin{equation}
\mathbf{M}= \langle\rv\rv^T \rangle,
\end{equation}
where the average is taken component-wise.
In the absence of nonlinearities, particular emittances are conserved. That
is in the case of coupling only the 6D emittance given by 
\begin{equation}
\eps_{6D}=\sqrt{\det \mathbf{M}}.
\end{equation}
In case of decoupled planes, the determinants of the diagonal blocks (the emittances of the respective plane) are conserved individually. 
They read
\beq
\eps_{x,n} =\frac{1}{m_e c} \sqrt{\det \mathbf{M_1}}\;,\;\;
\eps_{y,n} =\frac{1}{m_e c} \sqrt{\det \mathbf{M_2}}\;,\;\;
\eps_{z,n} =\frac{1}{e} \sqrt{\det \mathbf{M_3}}\;,\;\;
\label{defEmittance}
\eeq
in the usual units of m\,rad and eV\,s, respectively.
The analysis of emittance coupling by means of the eigen-emittances
\begin{equation}
\eps_{\mathrm{eig},i} = \mathrm{eigs} ( \mathbf{J M}),
\end{equation}
where $\mathbf{J}$ is the symplectic matrix, is also possible with our code, however beyond the scope of this paper.

\section{Continuous equations of motion}
\label{Analytic}
In order to address the continuous motion in DLA structures we employ positions and momentum as canonically conjugate variables in all directions. The transformation for the energy is $\Delta p_z= \Delta W /(\beta c)$. We address the flat and the tilted grating separately and assume for simplicity $|\ec_1|$ to be constant for all cells and $\arg(\ec_1)=0$.

\subsection{Flat grating}
\begin{figure}[t]
	\centering
	\includegraphics[width=0.8\linewidth]{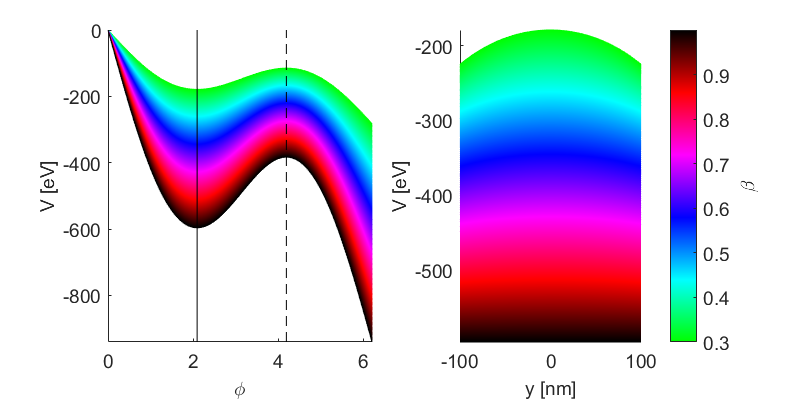}
    \caption{The potential (Eq.~\ref{Potential}) as function of $\phi$ and $y$ at the synchronous phase indicated by the solid vertical line. The longitudinally unstable fixed point (dashed line) flips the sign of the transverse potential. The color scale indicates the adiabatic change of the potential with $\beta$.}
    \label{Pic:Potential}
\end{figure}

Hamilton's equations can be written as
\begin{subequations}
\begin{align}
\dot x&=\frac{p_x}{m_e\gamma} \\
\dot p_x&=0\\
\dot y&=\frac{p_y}{m_e\gamma} \\
\dot p_y&=-q e_1 \frac{\lambda_{gz}}{2\pi}\frac{\omega}{\beta\gamma c} 
\sinh\left(\frac{\omega y}{\beta\gamma c}\right)
\sin\left(\frac{2\pi s}{\lambda_{gz}}\right)\\
\dot s&= \frac{\Delta p_z}{m_e\gamma^3}\\
\dot{\Delta p_z}&=q e_1 \left[
\cosh\left(\frac{\omega y}{\beta\gamma c}\right)
\cos\left(\frac{2\pi s}{\lambda_{gz}}\right)-\cos\phi_s\right].
\end{align}
\end{subequations}
Due to Eq.~\ref{PWtheorem} the force field is irrotational and can be derived from a potential as $\Fv=-\nabla' V$, where integration yields 
\begin{equation}
V=qe_1\left[ \frac{\lambda_{gz}}{2\pi}\cosh\left(\frac{\omega y}{\beta\gamma c}\right)
\sin\left(\frac{2\pi s}{\lambda_{gz}}\right)-s\cos\phi_s \right] .
\label{Potential}
\end{equation}
This potential and its adiabatic change with $\beta$ is illustrated in Fig.~\ref{Pic:Potential}.
The full 6D Hamiltonian reads
\begin{equation}
H=\frac{1}{2m_e\gamma}\left(p_x^2+p_y^2+\Delta P_z^2 \right)+V,
\end{equation}
where $\Delta p_z/\gamma$ was replaced with $\Delta P_z$. The coupled equations of motion are
\begin{subequations}
\begin{align}
\ddot x&= 0\\
\ddot y&= -\frac{q e_1}{m_e\gamma^2}
\sinh\left(\frac{\omega y}{\beta\gamma c}\right)
\sin\left(\frac{2\pi s}{\lambda_{gz}}\right)\\
\ddot s &= \frac{q e_1}{m_e\gamma^3}\left( 
\cosh\left(\frac{\omega y}{\beta\gamma c}\right)
\cos\left(\frac{2\pi s}{\lambda_{gz}}\right)-\cos\phi_s\right).
\end{align}
\end{subequations}
If the beam size is significantly smaller than the aperture ($y\ll \beta\gamma c/\omega$), the longitudinal equation decouples and becomes the ordinary differential equation of synchrotron motion. 
The transverse motion becomes linear in this case, however still dependent on the longitudinal motion via $\phi$. The equation of motion,
\begin{equation}
\ddot y= \frac{-q e_1 \omega}{m_e\gamma^3 \beta c}
\sin\left(\phi\right)y ,
\label{accdef}
\end{equation}
is Hill's equation, with the synchrotron angle being the focusing function. However there is a crucial difference to ordinary magnetic focusing channels. The focusing force scales as $\gamma^{-3}$ as expected for acceleration defocusing~\cite{Wangler2008RFAccelerators}, rather than with $\gamma^{-1}$ as would be expected for a magnetic quadrupole focusing channel.
The solution to Eq.~\ref{accdef} as function of $z$ for fixed $s=\lambda_{gz} \phi_s /2\pi$, i.e. when the bunch length is significantly shorter than the period length, is
\begin{equation}
y=y_0\exp\left(\sqrt{\frac{-q e_1 \omega}{m_e\gamma^3 \beta^3 c^3}
\sin\phi_s }z\right)
\end{equation}
and a synchronous particle with non-zero transverse offset is expected to grow to double transverse amplitude in 
\begin{equation}
L_2=\frac{\ln 2}{\sqrt{\frac{-q e_1 \omega}{m_e\gamma^3 \beta^3 c^3}
\sin\phi_s }}.
\end{equation}
For subrelativistic particles $L_2$ can reach down to a few micron. However, longer interaction lengths can be achieved by focusing the beam into the DLA structure externally.

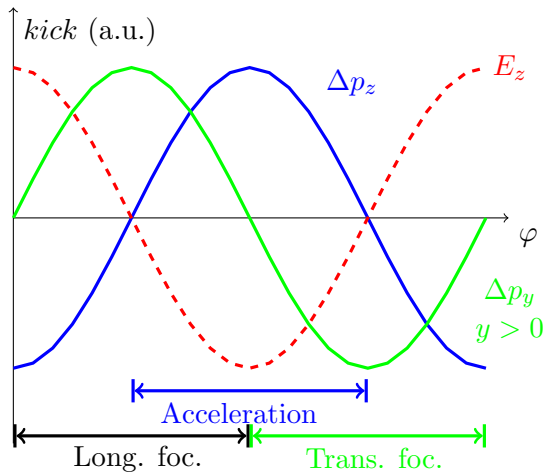
\begin{figure}[h]
\begin{tikzpicture}
  \draw[->] (0,0) -- (2*pi+0.3,0) node[below right] {$\phi$};
  \draw[->] (0,-3) -- (0,2.8) node[below right] {$kick$ (a.u.)};
  \draw [blue, very thick, domain=0:2*pi] plot ({\x}, {-2*cos(180/pi*\x)}) ;
  \draw [red,dashed, very thick, domain=0:2*pi] plot ({\x}, {2*cos(180/pi*\x)}) ;
  \node [red] at (6.6,2.0) {$E_z$};
    
  \draw [green, very thick, domain=0:2*pi] plot ({\x}, {2*sin(180/pi*\x)}) ;
  \node [green] at (6.6,-1.0) {$\Delta p_y$};
    \node [green] at (6.6,-1.5) {$y>0$};
    
  %\node [red] at (3.0,1.5) {$\Delta p_y$};
  \node [blue] at (4.5,1.8) {$\Delta p_z$};
	%\node [black!20!white] at (4,2) {Accel.};
	\draw[|<->|,black, very thick] (0,-2.9) -- (pi,-2.9);
  	\draw[|<->|,green, very thick] (pi,-2.9) -- (2*pi,-2.9);
	%\draw[|<->|,red,very thick] (pi,-3.0) -- (2*pi,-3.0);
	\node [black] at (4.8-pi,-3.2) {Long. foc.};
	\node [green] at (4.8,-3.2) {Trans. foc.};
	
   	\draw[|<->|,blue,very thick] (1/2*pi,-2.3) -- (3/2*pi,-2.3);
  	\node [blue] at (3.0,-2.6) {Acceleration};    
        
\end{tikzpicture}
\caption{Overview of electron acceleration and focusing properties for an x-invariant grating.}
\label{PhaseRelation}
\end{figure}

As shown in Fig.~\ref{PhaseRelation}, the phase ranges of longitudinal and transverse focusing are disjoint.
This is a consequence of Earnshaw's theorem~\cite{Earnshaw1842OnEther} which can be directly observed in Eq.~\ref{Potential}, i.e. V has no minima but only saddle points. 
Similarly as in Paul traps (see e.g.~\cite{Major2005ChargedTraps}), stable motion in both the y- and z-planes can only be achieved by rotating the saddle.
For an accelerator this means alternating the synchronous phase.
This so called Alternating-Phase-Focusing (APF) scheme has been developed for ion RF linacs already in the 1950' (e.g.~\cite{Fainberg1956AlternatingFocusing}) but later rejected in favor of the RFQ~\cite{Wangler2008RFAccelerators}.

For an adiabatic Hamiltonian and if stable orbits exist, 
a matched locally Gaussian distribution is given by
\begin{equation}
f = C e^{-H/\langle H \rangle}
\end{equation}
and a locally elliptic (Hofmann-Pedersen~\cite{Hofmann1979BunchesDistributions}) matched distribution is given by 
\begin{equation}
f = C \sqrt{H_{max} - H }.
\end{equation}
The normalization constant $C$ is determined by integration. Note that in the case of non-periodic motion $f$ will not be integrable.
Thus, we can only write a matched distribution for the longitudinal plane if $\phi_s\in [\pi/2,\pi]$ and for the transverse plane if $\phi_s\in [\pi,3/2\pi]$.

The Hamiltonian is not time independent, however its dependence on $\beta$ and $\gamma$ is adiabatic. Thus, if $\phi_s$ is changing at most adiabatically, the distribution will deform such that the emittance increase is bounded, i.e. also the emittance remains an adiabatic invariant.
First, we consider the longitudinal plane and linearized fields. For a given bunch length $\sigma_{\Delta s}$ the matched energy spread is 
\begin{equation}
\sigma_{\Delta W}=\frac{c_0}{\lambda_0} \sqrt {-2\pi\lambda_{gz} m_e \gamma^3 q e_1 \sin\phi_s } \sigma_{\Delta s}.
\label{EnergyMatching}
\end{equation}
For a slow change of the potential and filling the bucket only up to a small fraction, the phase space area given by $\pi \sigma_{\Delta\phi}\sigma_{\Delta W}$ is conserved. Moreover, using Eq.~\ref{EnergyMatching} a normalized bunch length and energy spread can be written as~\cite{Wangler2008RFAccelerators}
\begin{subequations}
\begin{align}
\sigma_{\Delta W,n}&=\frac{1}{\sqrt[4]{\beta^3\gamma^3}} \sigma_{\Delta_W} \\
\sigma_{\Delta\phi,n}&=\sqrt[4]{\beta^3\gamma^3} \sigma_{\Delta\phi}.
\end{align}
\end{subequations}
Accordingly, in position and momentum coordinates, this reads
\begin{subequations}
\begin{align}
\sigma_{\Delta P_{z,n}}&=\sqrt[4]{\frac{\beta}{\gamma}}\sigma_{\Delta P_{z}} 
\\
\sigma_{{\Delta s},n}&=\sqrt[4]{\frac{\gamma}{\beta}} \sigma_{\Delta s}.
\end{align}
\end{subequations}
Thus the adiabatic phase damping in DLAs behaves in the same way as in RF linacs. 

\begin{figure}[b]
	\centering
	\includegraphics[width=0.5\linewidth]{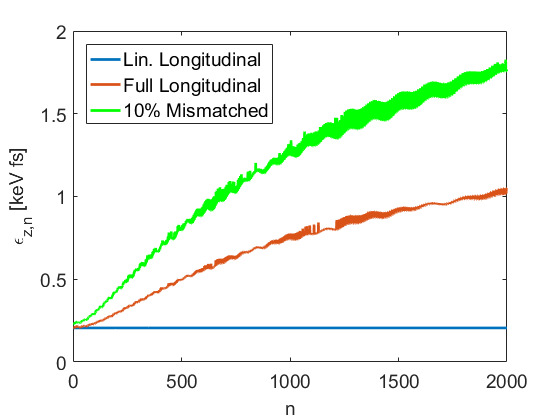}
    \caption{Longitudinal emittance evolution for a linearly matched Gaussian beam in linearized fields, with the full fields, and with 10\% excess energy spread. }
    \label{Longemittance}
\end{figure}

As a test of the code, we plot the long time evolution of the longitudinal emittance at zero transverse emittance for 3~different setups in  Fig.~\ref{Longemittance}. 
First, we consider a bunch matched according to Eq.~\ref{EnergyMatching} in linearized fields. As expected, the symplectic code preserves the emittance in linear fields. However, the linearly matched bunch shows emittance growth in the non-linear fields. Even stronger emittance increase is to be expected, when there is a mismatch of the bunch length and the energy spread (here we chose 10\% excess energy spread). The according result is obtained for the y-emittance when setting the synchronous phase into the transverse focusing regime and taking the longitudinal emittance as zero.

The effect of adiabatic damping also appears in DLAs in the transverse plane. 
The linearized transverse Hamiltonian reads
\begin{equation}
H_\perp = \frac{p_y^2}{2m_e\gamma} +\frac{q e_1 \omega \sin\phi_s}{2\beta \gamma^2 c} y^2
\end{equation}
and the matched momentum spread is 
\begin{equation}
\sigma_{p_y}=\sqrt{\frac{m_e q e_1 \omega \sin\phi_s}{\beta\gamma c}}\sigma_y.
\end{equation}
The area $\pi \sigma_y \sigma_{p_y}$ is conserved
and thus we can write normalized spreads as
\begin{subequations}
\begin{align}
\sigma_{y,n}&=\frac{1}{\sqrt[4]{\beta\gamma}}\sigma_{y}  \\
\sigma_{p_y,n}&=\sqrt[4]{\beta\gamma}\sigma_{p_y}.
\end{align}
\end{subequations}
One observes, that for increasing beam energy the transverse beam size increases, while the momentum spread decreases. This is in accordance with the potential becoming flatter in the transverse plane, while it becomes steeper in the longitudinal plane for increasing beam energy (cf. Fig.~\ref{Pic:Potential}).

\subsection{Tilted grating}
The Hamiltonian for the tilted grating is obtained by modifying the potential (Eq.~\ref{Potential}) as
\begin{equation}
V=q e_1 \left[\frac{\lambda_{gz}}{2\pi}\cosh\left(ik_y y\right)
\sin\left(\frac{2\pi s}{\lambda_{gz}}+\frac{2\pi x}{\lambda_{gx}}\right)-s\cos\phi_s\right].
\end{equation}
The coupled equations of motion are
\begin{subequations}
\begin{align}
\ddot x&= -\frac{q e_1}{m_e\gamma} \frac{\lambda_{gz}}{\lambda_{gx}}\cosh\left(ik_y y\right)
\cos\left(\frac{2\pi s}{\lambda_{gz}}+\frac{2\pi x}{\lambda_{gx}}\right) \\
\ddot y&= \frac{-ik_y \lambda_{gz} q e_1}{2\pi m_e\gamma}
\sinh\left(ik_y y\right)
\sin\left(\frac{2\pi s}{\lambda_gz}+\frac{2\pi x}{\lambda_{gx}}\right)\\
\ddot s &= \frac{q e_1}{m_e\gamma^3}\left[ 
\cosh\left(ik_y y\right)
\cos\left(\frac{2\pi s}{\lambda_gz}+\frac{2\pi x}{\lambda_{gx}}\right)-\cos\phi_s\right].
\end{align}
\end{subequations}
One can observe that a bunch which is not significantly shorter than the grating period is accelerated in both positive and negative $x$-direction dependent on $s$. Therefore, a coherent deflection can only be obtained for extremely short bunches. 
In the following we assume no net acceleration, i.e. $\phi_s=\pi/2$, and replace $s=\lambda_0\phi_s/2\pi+\Delta s$. 
Since tilted gratings are outlined for the generation of wiggler radiation we restrict ourselves to the ultra-relativistic case ($\beta\rightarrow 1$) here. 
From Eq.~\ref{ky} one finds $\pm ik_y=k_x=\omega/c \tan\alpha $, which simplifies the potential to 
\begin{equation}
V=qe_1 \frac{\lambda_0}{2\pi}\cosh\left(\frac{\omega \tan\alpha}{c}y\right)
\cos\left(\frac{\omega}{c}(\Delta s+x\tan\alpha)\right).
\end{equation}
The equations of motion become
\begin{subequations}
\begin{align}
\ddot x&= \frac{q e_1}{m_e\gamma} \tan\alpha
\cosh\left(\frac{\omega \tan\alpha}{c}y\right)
\sin\left(\frac{\omega}{c}(\Delta s+x\tan\alpha)\right) \\
\ddot y&=  \frac{-ik_y \lambda_{gz} q e_1}{2\pi m_e\gamma}
\sinh\left(\frac{\omega \tan\alpha}{c}y\right)
\cos\left(\frac{\omega}{c}(\Delta s+x\tan\alpha)\right)\\
\ddot{\Delta s} &= \frac{q e_1}{m_e\gamma^3} 
\cosh\left(\frac{\omega \tan\alpha}{c}y\right)
\sin\left(\frac{\omega}{c}(\Delta s+x\tan\alpha)\right).
\end{align}
\end{subequations}
Injecting the beam with an offset $x_0 \ll \lambda_0/(4\tan\alpha)$
results in a coherent oscillation around the $x$-axis with the longitudinal period
\begin{equation}
\lambda_u= \frac{2\pi c}{\sqrt{\frac{-2 \pi q e_1}{m_e\gamma \lambda_0} }\tan\alpha}
\label{lambda_u}
\end{equation}
for a particle with $\Delta s=y=0$.
In linearized fields, the oscillation amplitude is arbitrary.
However, in the nonlinear fields, additionally to the longitudinal plane, we find 'buckets' with a distance $\lambda_{gx}$ also in the $x$-plane. These buckets split a ribbon beam which is large in x-direction into multiple beamlets, where the momentum spread acceptance is maximum at zero and vanishes at $\lambda_{gx}/2$, where  
 integer multiples of $\lambda_{gx}$ can be added.

\section{Applications}
\label{Applications}
We apply our approach to similar experimental parameters as for the subrelativistic experiments at FAU Erlangen~\cite{Breuer2013Laser-BasedStructure} and the relativistic experiments at SLAC~\cite{Peralta2013DemonstrationMicrostructure.,Wootton2016DemonstrationPulses}. Although the structures are idealized, the results are qualitatively recovered. As a next step, we show modifications and idealizations of the beam parameters, which outline the way to a microchip accelerator.

\subsection{Subrelativistic Acceleration}
A subrelativistic DLA structure needs to be chirped in order to always fulfill the synchronicity condition~\ref{phasesynchronicity} for the synchronous particle. 
\begin{figure}[h]
	\centering
	\includegraphics[width=0.47\linewidth]{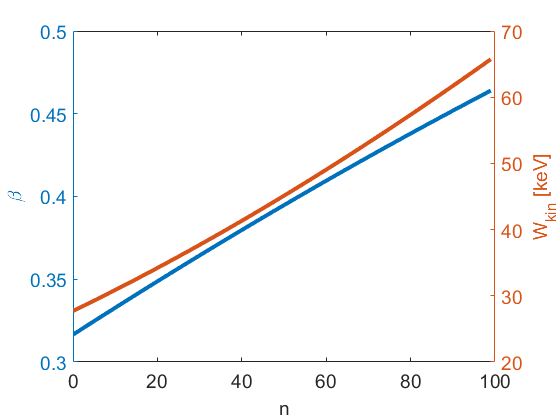}
    \caption{Acceleration ramp according to Eqs.~\ref{Ramp1}~and~\ref{Ramp2}.}
    \label{Ramp}
\end{figure}
The proper chirp for each cell and the synchronous velocity are obtained by iterating the two equations 
\begin{equation}
\Delta z^{(n+1)}=\frac{q e_1 \lambda_0^2 \cos \phi_s^{(n)}}{m_e c^2}  \sqrt{1-\beta^{{(n)}^2}}^3
\end{equation}
\begin{equation}
\beta^{(n+1)}=\beta^{(n)}+\frac{\Delta z^{(n+1)}}{\lambda_0}.
\label{Ramp1}
\end{equation}
The cell length and synchronous energies are
\begin{equation}
\lambda_g^{(n)}=\lambda_{g0} + \sum\limits_{j=2}^n  \Delta z^{(j-1)}
\label{celllength}
\end{equation}
\begin{equation}
W_0^{(n)}= W_\mathrm{init} +q e_1  \sum\limits_{j=2}^n \cos\phi_s^{(j-1)}  \lambda_g^{(j-1)}.
\label{Ramp2}
\end{equation}
In the following simulations, we assume that the construction of the grating was made such that the cells always fulfill Eq.~\ref{celllength}.
We start with very low energy electrons $W_\mathrm{kin}=27.7$~keV, i.e. $\beta=0.3165$.

\begin{figure}[t]
	\centering
	\includegraphics[width=\linewidth]{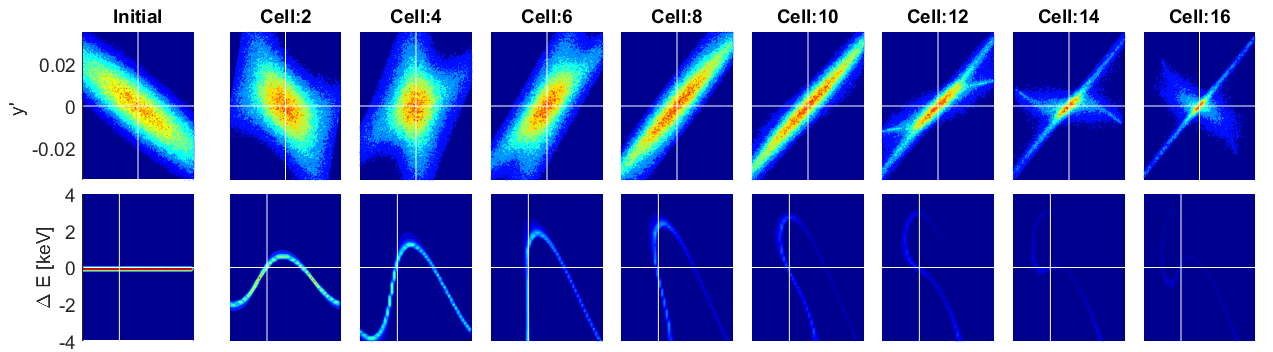}
        \caption{Longitudinal (top) and transverse (bottom) phase space for 16~cells of the chirped grating with an initially unbunched beam, focused into the structure. The axes are $100 \mathrm{nm}<y<100 \mathrm{nm}$ and $0<\phi<2\pi$. The color represents the phase space density, where the initial plots are normalized to their maximum and all other plots are normalized to the respective maxima of the second column.}
    \label{SubrelCoasting}
\end{figure}
\begin{figure}[b]
	\centering
    \includegraphics[width=0.47\linewidth]{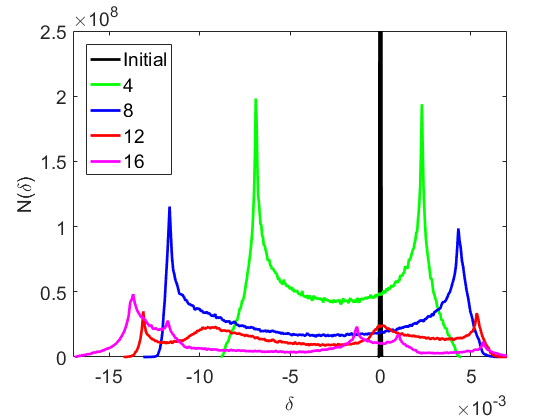}
	\includegraphics[width=0.47\linewidth]{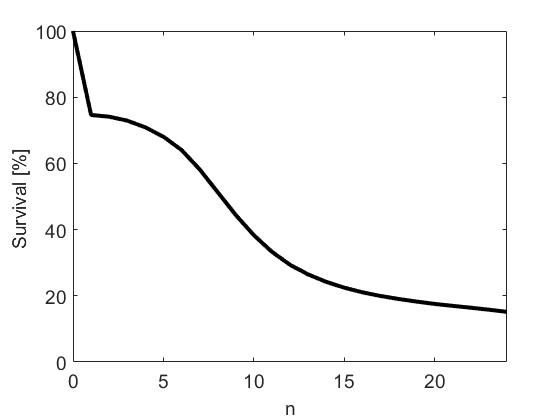}
        \caption{Energy spectra and particle survival rate for the unbunched beam.}
    \label{SubrelSurvivalCoasting}
\end{figure}
\begin{figure}[t]
	\centering
	\includegraphics[width=\linewidth]{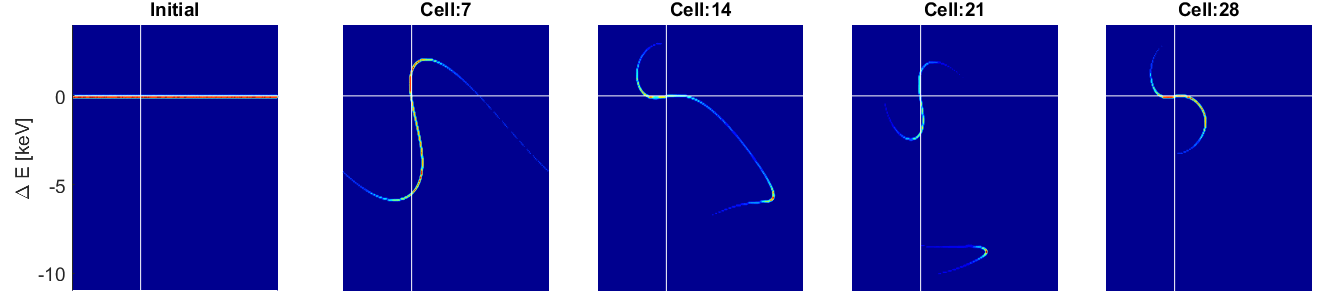}
        \caption{Longitudinal phase space evolution for zero transverse emittance. The vertical axis is $0<\phi<2\pi$.}
    \label{SubrelCoastingLongOnly}
\end{figure}

For $\lambda_0=1.96\;\mu$m the initial grating period is 620~nm. For simplicity we assume ${e_1=1}$~GV/m with zero phase for all cells. Aiming for a gradient of 500~MeV/m, the synchronous phase has to be 120 degree. The ramp according to these parameters is depicted in Fig.~\ref{Ramp}. Since the electron bunches in the experiments are significantly longer than the grating period, we look at initially unbunched beams with $\sigma_E=10$~eV. 

In free space, the full Rayleigh length (cf. App.~\ref{Rayleigh}) of the beam with assumed geometric emittance 1~nm at the aperture of $A=200$~nm is $L_R=A^2/(4\eps_y)= 10~\mu$m, i.e. about 16~cells. This requires the optimal initial focusing angle of 20\,mrad. However, as shown in Fig.~\ref{SubrelCoasting}, in the presence of strong acceleration defocusing forces, the waist appears earlier, i.e. in cell 4.

The particle loss and the energy spectrum are plotted in Fig.~\ref{SubrelSurvivalCoasting}. 
The spectrum shows clearly that only a fraction of the particles is trapped in the bucket, the particles with $\delta\approx -13\cdot10^{-3}$ are lost, although they do not hit the aperture. The physical loss of particles happens when they reach the aperture in y-direction ($\pm 10$\,nm).
The longitudinal bucket capture process is illustrated more clearly for zero transverse emittance in Fig.~\ref{SubrelCoastingLongOnly}, where no transverse losses appear and a full synchrotron period is displayed.

 \begin{figure}[b]
	\centering
	\includegraphics[width=\linewidth]{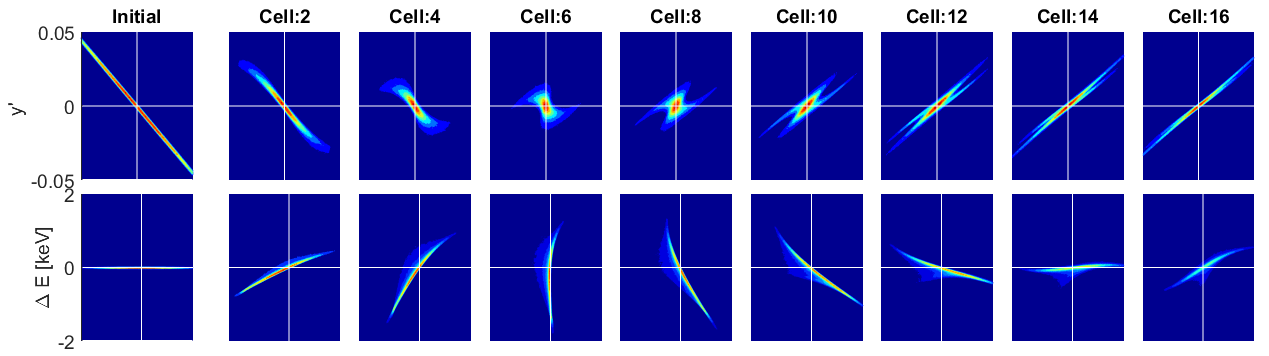}
        \caption{Longitudinal (bottom) and transverse (top) phase space for 16~cells of the chirped grating with the beam initially focused into the structure. The axes are $100 \mathrm{nm}<y<100 \mathrm{nm}$ and $1.5<\phi<2.6$. Again, the color represents the phase space density, normalized to the second column.}
    \label{SubrelBunched}
\end{figure}
\begin{figure}[t]
	\centering	
    \includegraphics[width=0.47\linewidth]{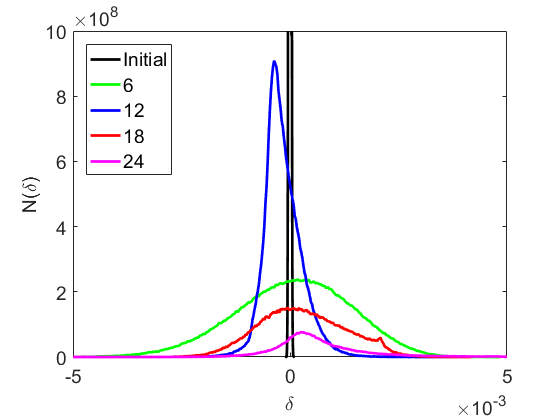}
	\includegraphics[width=0.47\linewidth]{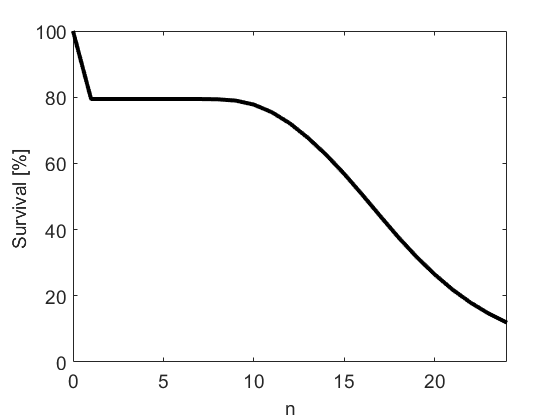}
        \caption{Energy spectra and particle survival rate for a short low energy bunch.}
    \label{SubrelBunchedSurvival}
\end{figure}
As next step, we take a bunched beam with $\sigma_z=30$\,nm and a reduced transverse emittance of $\eps_y=0.1$\,nm. As shown in Fig.~\ref{SubrelBunched}, the waist appears approximately at cell 7, when the beam is strongly focused initially with 45\,mrad. Without the acceleration defocusing, the Rayleigh length would be 100\,$\mu$m at an initial focusing angle of 2\,mrad. 
The bunch has initially again an energy spread of 10\,eV, which is significantly smaller than the matched energy spread. Thus a coherent quadrupole oscillation appears, which is also visible in the energy spectrum in Fig.~\ref{SubrelBunchedSurvival}.

\subsection{Relativistic Acceleration}
In this example we take the full aperture to be $A=800$~nm, $\lambda_g=\lambda_0=1.96\,\mu$m, $|\ec_1|=1$~GV/m, and the number of grating cells is 100. Taking the reference particle on-crest, i.e. $\phi_s=\pi$, the design ramp is linear with a slope of 1 GeV/m.
\begin{figure}[b]
	\centering
	\includegraphics[width=\linewidth]{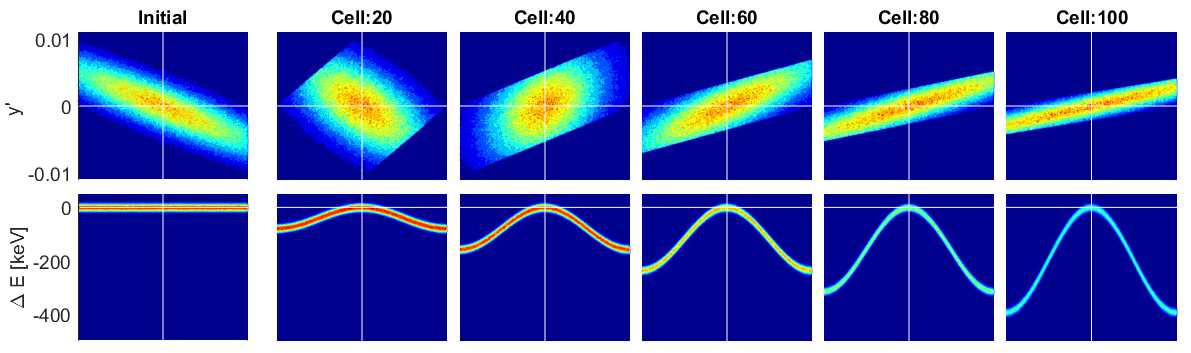}
    \caption{Evolution of the y and z phase spaces. The axes are $400 \mathrm{nm}<y<400 \mathrm{nm}$ and $0<\phi<2\pi$. Again, the color represents the phase space density, normalized to the second column.}
    \label{RelPhaseSpace}
\end{figure}
The incident electron beam has a bunch length significantly larger than the grating period and is again assumed as unbunched. The kinetic energy is 60~MeV and the spread is $\sigma_W=10$~keV. The spot size is taken as $\sigma_y=400$~nm and the geometric emittance $\eps_y=1$~nm. The full Rayleigh range is thus $L_R=A^2/(4\eps_y)=160\;\mu$m, which is practically achieved when the beam is focused with 5~mrad into the structure.
Figure \ref{RelPhaseSpace} shows the evolution of the y and z phase spaces. 
\begin{figure}[t]
	\centering
	\includegraphics[width=0.49\linewidth]{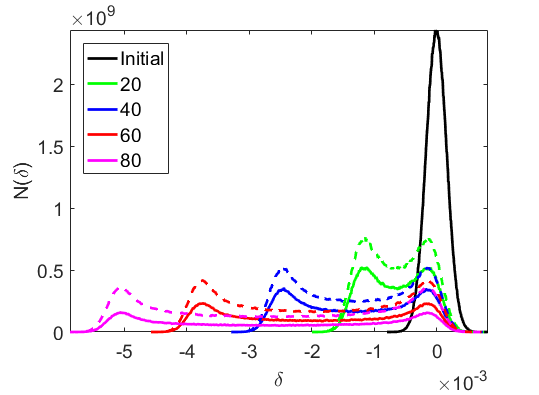}
    	\includegraphics[width=0.47\linewidth]{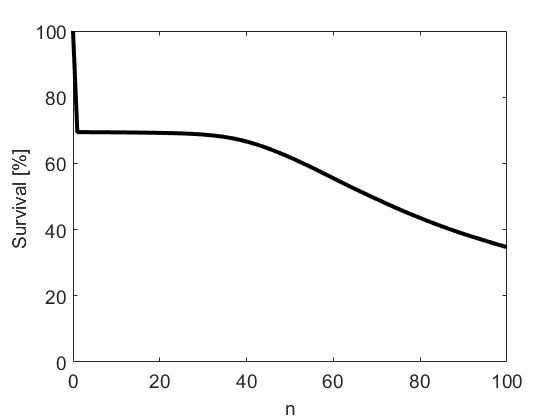}
        \caption{Energy spectra with 1\,nm (solid lines) and zero (dashed lines) transverse emittance normalized to the initial number of particles $10^6$ (left) and particle survival rate (right).}
    \label{EnergySpectrum}
\end{figure}
The particle loss is monitored in Fig.~\ref{EnergySpectrum}, where the first jump is again the loss at the initial aperture. The plateau is within the Rayleigh range, however already before the end of the Rayleigh range the particles with excess momentum are being lost. Exactly at the Rayleigh range (after cell 81), the particle, that was initially the intersection of the ellipse diagonal with the aperture, is lost. The acceleration defocusing plays only a minor role for highly relativistic beams, i.e. the Rayleigh range is not significantly shortened. Figure~\ref{EnergySpectrum} shows also the energy spectrum which becomes broader along the grating. This is due to particles being accelerated and decelerated according to their phase. Such spectra were also practically measured in~\cite{Peralta2013DemonstrationMicrostructure.,Wootton2016DemonstrationPulses}. The dashed lines in the plot show the same spectra in the case of zero transverse emittance, where also no loss on the aperture occurs.

\subsection{Dynamics in Tilted Gratings} 
\begin{figure}[h]
	\centering
	\includegraphics[width=\linewidth]{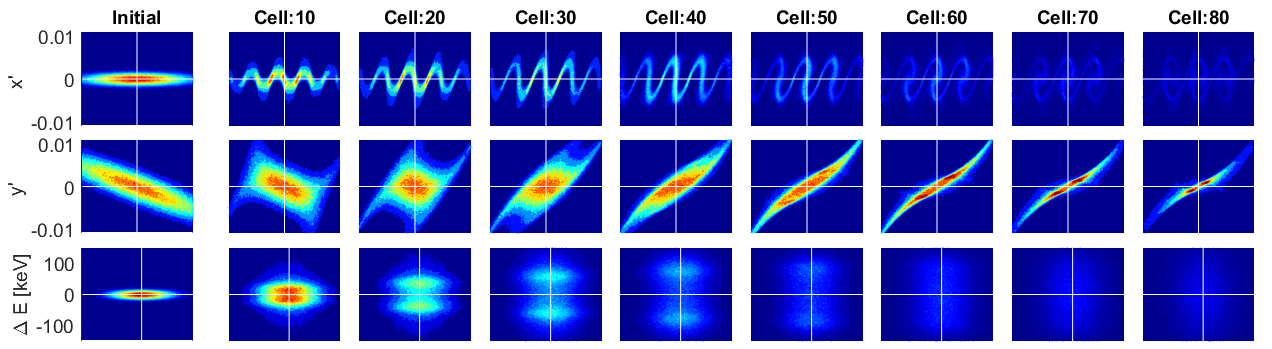}
        \caption{Horizontal, vertical, and longitudinal phase space projection for every tenth cell. The axes are $-1.5\;\mu \mathrm{m}<x<1.5\;\mu \mathrm{m}$, $-0.4\;\mu \mathrm{m}<y<0.4\;\mu \mathrm{m}$ and $1.3<\phi<1.8$. Again, the color represents the phase space density, normalized to the second column.}
    \label{TiltedPhaseSpace}
\end{figure}
\begin{figure}[h]
	\centering
	\includegraphics[width=0.47\linewidth]{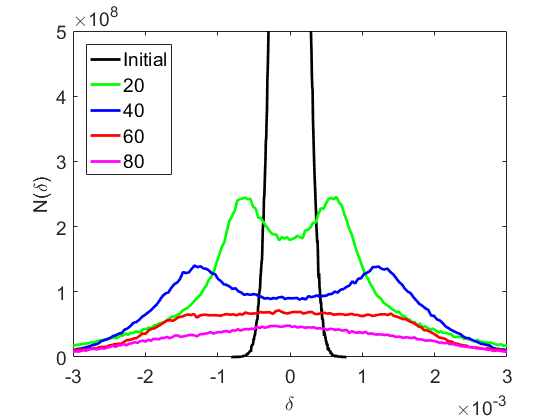}
        	\includegraphics[width=0.47\linewidth]{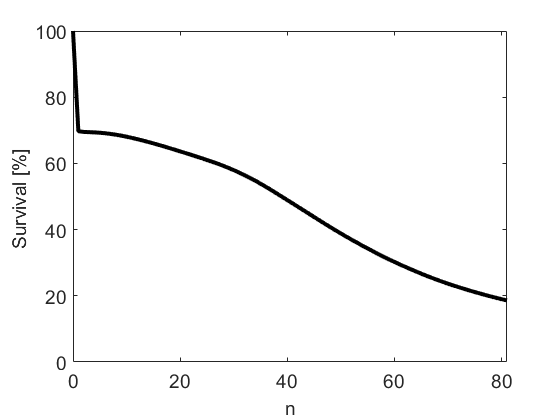}
      \caption{Evolution of the energy spectrum and particle loss.}
    \label{TiltedESpec}
\end{figure}
\begin{figure}[h]
	\centering
	\includegraphics[width=\linewidth]{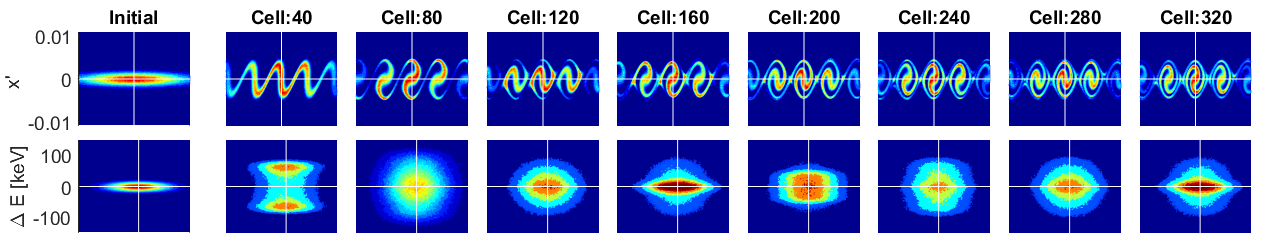}
        \caption{Horizontal and longitudinal phase space for every fortieth cell for $\eps_y=0$. The axes are $-1.5\;\mu \mathrm{m}<x<1.5\;\mu \mathrm{m}$ and $1.3<\phi<1.8$. Two periods of transverse oscillations are visible (cf. Eq.~\ref{lambda_u}).}
    \label{TiltedPhaseSpaceZeroEpsY}
\end{figure}
Finally, we address the tilted grating with the same laser parameters and a bunched electron beam with parameters $\eps_x=\eps_y=1$\,nm, $\sigma_x=1\,\mu$m, $\sigma_y=0.4\,\mu$m, $\sigma_z=30$\,nm, $\sigma_W=10$\,keV and a focusing angle of 5\,mrad in the y-direction. The grating tilt angle is 70~degrees and again $|\ec_1|=1$~GV/m.
Figure~\ref{TiltedPhaseSpace} shows the evolution of the phase space in all three planes. Evaluating Eq.~\ref{lambda_u}, one finds $\lambda_u\approx 160\lambda_0$, i.e. half an oscillation period in the x-direction in the displayed 80 grating cells.

As visible in Fig.~\ref{TiltedPhaseSpace}, the horizontal and longitudinal phase spaces are correlated. The projections of the energy spectrum can be seen in Fig.~\ref{TiltedESpec} together with the particle loss, which takes place at the physical aperture in y-direction at $\pm 400$~nm.
Unlike the straight grating with relativistic particles, the tilted grating creates a defocusing force in the y-direction which significantly decreases the Rayleigh range. The energy spread shows a breathing mode, similar to the quadrupole modes in the synchrotron motion. However, since the synchrotron motion is practically frozen due to the high $\gamma$, this mode arises entirely due to the correlation with the x-plane. Excluding the defocusing by setting $\eps_y=0$, two coherent oscillation periods are displayed in Fig.~\ref{TiltedPhaseSpaceZeroEpsY}. 

\section{Conclusion and Outlook}
\label{Conclusion}

The laser fields in a periodic DLA grating can be represented by spatial Fourier harmonics, where only the resonant harmonic, which fulfills the Wideroe condition, provides a first order net kick. Exploiting this property, we showed that the entire 6D beam dynamics without collective effects can be modeled by applying kicks in all spatial directions once per grating period. These kicks are not independent, but analytically connected by the Panofsky-Wenzel theorem. If the structure is not periodic, but slightly chirped (quasi-periodic), our approach is still applicable. However, fringe fields at the end of the structure are neglected. 

As an example we introduced a novel Bragg-reflection based grating structure, which shows a  particularly high first harmonic. The structure is discussed in~\cite{Niedermayer2017DesigningChip} and \cite{Egenolf2017SimulationDomain} in more detail. Here we restrict ourselves in representing the grating structures by the one resonant Fourier coefficient, i.e. one complex number for each grating cell. 

We also showed that our tracking approach still works for tilted gratings that have been proposed for beam deflection or optical undulators. However, in the case of curved gratings, which have been proposed for focusing, the fields cannot be determined analytically, since the decay constant is not uniform. In order to still use our tracking algorithm, the longitudinal kicks must be provided numerically for each pair of transverse coordinates.

Additionally to our fast, symplectic tracking approach, we also derived the Hamiltonian for the single particle motion in DLA structures. This allows analytical approaches to the 6D nonlinear and coupled equations of motion in DLA structures. In  the case of constant synchronous phase, the longitudinal beam dynamics is identical to the one for conventional drift tube linacs. However, for longitudinally stable buckets, the transverse fields are always defocusing. Due to the high gradients in DLA, this strong defocusing cannot be compensated by ordinary means as magnets, which is particularly critical at low electron beam energy. At relativistic energies the full Rayleigh range of an externally focused beam can be reached, however this is also only in the range of several hundred microns. 

In future a focusing scheme for DLA needs to be developed. One candidate is the proposed higher order harmonic focusing~\cite{Naranjo2012StableHarmonics}. In order to simulate this with our code, the additional harmonic kicks would have to be implemented. The other candidate is Alternating Phase Focusing (APF), which can be directly approached with our code. In this scheme the synchronous phase is alternated between longitudinally stable and unstable ranges, similarly as a FODO cell, but instead of x-y, rather in the y-z planes. 

We plan to achieve such phase jumps by inserting drift sections as already outlined in Fig.~\ref{APF}. Other options are to modify the accelerating Fourier coefficient in each cell, e.g. by phase masking within the structure or by active phase control of individual parts of the laser pulse. In general, we believe that this paper gives a beam dynamics foundation on which DLA structures providing stable long distance beam transport schemes can be developed.

\begin{acknowledgments}
The authors wish to thank Ingo Hofmann for proofreading the manuscript.
This work is funded by the Gordon and Betty Moore Foundation (Grant GBMF4744 to Stanford) and the German Federal Ministry of Education and Research (Grant FKZ:05K16RDB).
\end{acknowledgments}

\appendix
\section{Rayleigh range for light and particle beams}
\label{Rayleigh}
The Rayleigh range for a particle beam can be defined in the same way as as for a light beam. The envelope of an externally focused beam is 
\begin{equation}
w=w_0\sqrt{1+\left(\frac{z}{z_0}\right)^2}.
\end{equation}
Inserting into the envelope equation
\begin{equation}
w''=w^{-3} 
\end{equation}
results in $w_0=\sqrt{z_0}$. The beam size is given by $a(z)=\sqrt{\eps}w(z)$ and thus the Rayleigh length is $z_0=a_0^2/\eps$, where $a_0$ is the beam size at the waist. The beam size at the Rayleigh length, where it is limited by the aperture, is $a_1=\sqrt{2}a_0$. The full Rayleigh range $L_R=2z_0$ as function of the full aperture $A=2a_1$ is thus
\begin{equation}
L_R=\frac{A^2}{4\eps}.
\end{equation}
The same result is obtained for a light beam, where the emittance is identified with the wavelength, i.e. $\eps_\mathrm{light}=\lambda_0/\pi$.

\bibliography{Mendeley}
\end{document}

%% file: TEDmakros.tex
\newcommand{\beq}{\begin{equation}}
\newcommand{\eeq}{\end{equation}}

\newcommand{\eps}{\varepsilon}

\renewcommand{\rho}{\varrho}
\renewcommand{\theta}{\vartheta}
\renewcommand{\phi}{\varphi}

\newcommand{\ex}{\ensuremath{ \vec{e}_x} }

\newcommand{\ey}{\ensuremath{ \vec{e}_y} }

\newcommand{\Bv}{\ensuremath{ \vec{B} }}

\newcommand{\Fv}{\ensuremath{ \vec{F} }}

\newcommand{\fvc}{\ensuremath{\protect \underline{\vec{f}} }}

\newcommand{\ec}{\ensuremath{ \underline{e} }}
\newcommand{\Ec}{\ensuremath{ \underline{E} }}

\newcommand{\rv}{\ensuremath{ \vec{r} }}

\newcommand{\pv}{\ensuremath{ \vec{p} }}

\newcommand{\ds}{\ensuremath{ \mathrm{d}s}}

\newcommand{\dz}{\ensuremath{ \mathrm{d}z}}
\newcommand{\dt}{\ensuremath{ \mathrm{d}t}}

\renewcommand{\Re}{{\rm Re} \,}
\renewcommand{\Im}{{\rm Im} \,}

\newcommand{\wegdamit}[1]{} %% auskommentieren

\newlength{\lwveryfine}   
\newlength{\lwfine}   
\newlength{\lwnormal} 
\newlength{\lwthick}  
\newlength{\lwverythick}  